\documentstyle[12pt,epsfig]{article}
\title{
\vspace{-36pt}
{\normalsize \begin{flushright}SUSX-TH-96-020\\
{\tt astro-ph/9612135}\\
\end{flushright}}
\vspace{1 cm}
{\bf Scaling and Small Scale Structure in Cosmic String Networks}}
\author{
Graham R. Vincent$^{(a)}$\thanks{E-mail: {\tt 
g.r.vincent@sussex.ac.uk}}\\
Mark Hindmarsh$^{(a)}$\thanks{E-mail: {\tt 
m.b.hindmarsh@sussex.ac.uk}}\\
Mairi Sakellariadou$^{(b)}$\thanks{E-mail: {\tt 
mairi@karystos.unige.ch}}}
\date{\today}

\begin{document}

\maketitle
\vspace{-30pt}
\begin{center}
{\normalsize {\it
$^{(a)}$Centre for Theoretical Physics\\ 
University of Sussex\\
Brighton BN1 9QH\\
UK\\[6pt]
$^{(b)}$D\'epartement de Physique Th\'eorique\\
Universit\'e de Gen\`eve\\
Quai Ernest-Ansermet 24\\
CH-1211 Gen\`eve 4\\
Switzerland}
}
\end{center}

\begin{abstract}
We examine the scaling properties
of an evolving network of strings
in Minkowski spacetime and study the evolution of
length scales in terms of a 3-scale
model proposed by Austin, Copeland and Kibble (ACK).
We find good qualitative and some quantitative
agreement between the model and our simulations.
We also investigate small-scale structure by altering
the minimum allowed size for loop production $E_c$.
Certain quantities depend significantly on this parameter:
for example the scaling density can vary by a factor of two
or more with increasing $E_c$.
Small-scale structure as defined by ACK disappears if no 
restrictions are placed
on loop production, and the fractal dimension of the string changes
smoothly from 2 to 1 as the resolution scale is decreased.
Loops are nearly all produced at the lattice cut-off.
We suggest that the lattice cut-off should be interpreted
as corresponding to the string width, and that in a real network 
loops
are actually produced with this size. This leads to a radically
different string scenario, with particle production rather than 
gravitational
radiation being the dominant mode of energy dissipation.
At the very least, a better understanding of the
discretisation effects in all simulations of cosmic strings is
called for.
\end{abstract}

\section {Introduction}

Cosmic strings formed during a cosmological phase transition
at an energy scale of about $10^{16}$ GeV are sufficiently massive
to seed structure formation. However, calculations of the
gravitational effects from strings are hampered by
uncertainty over the statistics of the evolving network 
\cite{HindKib94,ShelVil}.

From lattice simulations of string formation \cite{Lattice},
a picture has developed of a string network at formation consisting
of a scale-invariant distribution of closed loops
together with a percentage of infinite brownian strings
crossing the Universe. The subsequent evolution would consist 
of the growth of the step length of the brownian infinite string 
network
as the strings attempt to straighten out, and
energy loss via loop production during reconnection.
 
Early analytic work identified the key 
property of {\it scaling} where at least the gross properties of the 
network 
can be characterised by a length scale, roughly the persistence 
length or
the interstring distance $\xi$, which grows with the horizon 
\cite{Kibble85}.
This result was supported by subsequent numerical work
\cite{AlbTur8589}. However, further investigation revealed 
dynamical
processes, including loop
production, at scales much smaller than $\xi$ 
\cite{FRWCodes,SakVil90}. In response,
Austin {\it et al} developed a model describing the network
in terms of three scales \cite{ACK} (hereafter ACK). 
These scales are: the usual energy density scale $\xi$,
a correlation length $\bar\xi$ along the string
and a scale $\zeta$ relating to local structure on the string.
It seemed likely from the ACK model that $\xi$ and $\bar\xi$ would 
scale with $\zeta$ growing slowly, if at all, until gravitational 
radiation
effects became important when $\zeta/\xi\approx 10^{-4}$
\cite{GravInf}.  Progress 
has also
been made using alternative analytical approaches in 
\cite{Hind96,Martins96}.

Understanding how the network scales is important
because it simplifies the process of building a model representing
the evolving network. Predictions of the 
power in the anisotropies in the Microwave Background require
the two-time correlation functions for various components of the
string energy-momentum tensor evolving over $O(10^3-10^4)$ expansion 
times.
As present simulations cannot achieve this range, constructing a 
model
is necessary. We presented a simple model of the two-time 
correlation 
functions in a non-expanding background in \cite{VHS}.

In this paper we present the results of a numerical study of
evolving strings in a Minkowski space-time. We demonstrate 
the scaling behaviour of correlation functions of ${\bf p}$ and 
${\bf q}$ vectors
(dynamical quantities which are linear combinations of tangent and
velocity vectors) along the string. These functions may be used to
complement analytical studies \cite{ACK} and
also in existing formalism to predict small-scale anisotropy in
the cosmic microwave background $\cite{Hind93}$. The largest length 
scale
associated with these correlations, $\bar\xi$, and the
interstring distance $\xi$ are both significantly smaller than the
causal horizon which suggests that the effects of the expanding
background on the correlation functions are small. Following
Coulson {\em et\, al\,} \cite{Coul94}, we think of the Minkowski
network of strings as modelling more realistic strings, by mapping
Minkowski time to conformal time and Minkowski space to comoving 
coordinates. 

As a test for the ACK model, we examine the rate equations
for the evolving length scales developed in ACK as they apply
to a Minkowski space-time with no gravitational radiation, and 
compare the
results with our simulations. Such a comparison is problematic
because there are a number of parameters which are difficult to
calculate directly. In particular, the parameter $k$ which controls
the behaviour of $\zeta$ is hard to measure. We have showed good
quantitative agreement between the ACK model and our simulations for 
the approach to 
scaling of $\xi$, and qualitative agreement for $\zeta$.
 
A major problem is understanding the relationship between 
small-scale
structure along the string and loop production.
Bennett and Bouchet \cite{FRWCodes} provide evidence that strings 
possess a fractal substructure arising from kink production during 
reconnection.
This ``intermediate'' fractal, constant at fixed time, spreads over 
a larger range of scales
as the network evolves and by the end of the simulation covers a 
range
between the resolution scale and the persistence length $\bar\xi$.
Most of the loop production occurs at scales near the resolution 
scale.
Shellard and Allen \cite{FRWCodes}
also observe loop production near the scale of resolution and an 
``intermediate'' fractal
in the long string, although they attribute the fractal
to the effect of initial conditions and a restrictive dynamical 
range.

The build up of small-scale structure is allowed by 
the Austin {\em et\, al\,} analysis: their original guesses
for the model parameters yield the increase in
small-scale structure observed in the expanding Universe codes.
However, it is not well known that their model can also predict 
that the small-scale structure (as measured by $\zeta$)
is absent. 

Indeed, we find that if there are no restrictions on loop 
production,
then small-scale structure will disappear: all three length scales
defined in ACK will have the same magnitude. We also find that the
region of constant fractal dimension, which we observe if we
artificially constrain loop production, disappears. This
has interesting implications for the energy loss mechanism
from cosmic strings.
 
The standard scenario is that any build up of small-scale structure
will only be checked by the back reaction from the string's own
gravitational field, which only becomes effective when $\zeta/\xi$
is small (of order $10^{-4}$) \cite{GravInf}. Loop production will
then 
grow with the horizon, albeit at a scale much smaller than $\xi$.
Although much smaller than $\xi$, these loops are vastly bigger than
the string width and will be topologically stable, losing energy 
through 
gravitational radiation.

If there is no small-scale (as measured by $\zeta$)
and loop production occurs at the smallest
possible scale - presumably the string width -
then energy loss will be dominated by production
of GUT quanta \cite{Srednicki87}. Furthermore, particle production
from cuspy regions may occur {\em without} any conventional loop 
production
at all. This will obviously allow the string scenario to avoid 
any gravity wave bounds \cite{HindKib94}, but the implications for 
the decay products
of the string quanta are less clear.  Recent calculations of the 
flux
of high energy decay products from string quanta have been compared 
with observed fluxes
to put an upper bound on the proportion of string energy going into 
particles
\cite{PartBound}. In the standard scenario these bounds can be 
avoided \cite{Gill94}.
However, if energy loss is dominated by particle production such 
bounds will constrain
the GUT physics if the GUT scale string scenario is to survive.

\section{Algorithm and Code}

In Minkowski space the string equations of motion are
\begin{equation}
{\bf X}'' = \ddot {\bf X},
\label {Waveequation}
\end{equation}
where 
$\dot{\bf X}={\partial{\bf X}/ \partial\eta}$ and 
${\bf X}'={\partial{\bf X}/ \partial{s}}$; 
$\eta$ 
is time
and ${s}$ is a space-like parameter along the string.
The spatially flat FRW metric is conformal to the Minkowski metric
and we regard the Minkowski time as conformal time, in the limit
that the expansion of the Universe goes to zero.

${\bf X}={\bf X}({s},\eta)$ is a position three-vector which 
satisfies the constraints
\begin{equation}
{\bf X}'\cdot\dot {\bf X}=0,
\label {GaugeCond1}
\end{equation}
\begin{equation}
{\bf X}'^2+\dot {\bf X}^2=1.  
\label {GaugeCond2}
\end{equation}
The first constraint ensures no tangential velocities,
and (\ref {GaugeCond2}) ensures that the energy of
a segment of string is proportional to its length when
measured in units of $s$.

Using a development of a code written by one of us  previously 
\cite{SakVil90,SakVil88}, we solve the wave Eq. (\ref{Waveequation})
with the Smith-Vilenkin algorithm \cite {SmithVil:alg}. This 
algorithm
uses the exact finite difference solution to (\ref {Waveequation}),
\begin{equation}
{\bf X}({s},\eta+\delta) = {\bf X}({s}+\delta,\eta) + {\bf 
X}({s}-\delta,\eta) - {\bf X}({s},\eta-\delta).
\label {SmithVil}
\end{equation}
If the string points are initially defined on the sites of a cubic 
lattice $(N\delta)^3$,
then (\ref {SmithVil}) ensures that they remain on the lattice at 
time steps of $\delta$. The discretised gauge conditions require 
that for
these initial conditions, string links are restricted to three 
types:
stationary (${|{\bf \dot X}|}=0$,\,\,${| {\bf X'}|}=1$), diagonal
(${|{\bf \dot X}|}=1/\sqrt{2}$,\,\,${| {\bf X'}|}=1/\sqrt{2}$) and 
cusps 
(${|{\bf \dot X}|}=1$,\,\,${| {\bf X'}|}=0$).

Because the string points lie on the sites of the lattice, 
identifying crossing events is simple.
When two strings cross, they intercommute with a probability 
which is set to $P_I$. For most of this work we set $P_I=1$,
subject to the condition
that reconnection does not create a loop smaller than
the threshold $E_c$. Loops with energy greater than or equal
to a threshold value
of $E_c$ are allowed to leave the network,
while reconnections are forbidden for loops with energy equal to 
$E_c$. Forbidding reconnections allows energy to leave the network
fairly efficiently; otherwise it takes much longer for the effect
of the initial conditions to wear off. This feature may also model
more realistic networks as, in an FRW Universe, small loops will
decouple from the expansion and are highly unlikely to reconnect.

An interesting value for $E_c$ is the minimum segment length 
$2\delta$. The loops produced in this case are ``cusps'': confined
to one lattice site and travelling at the speed of light. The
cutoff then becomes a lattice cutoff. 

The ability to alter $E_c$ allows us a certain amount of control
over small-scale structure and related quantities. For example,
as $E_c$ is increased from $2\delta$ to $8\delta$, the rms velocity
of a segment along the string increases from $0.36$ (very close to
the matter era value in Bennett and Bouchet) to $0.46$ (close
to the radiation era value).
This corresponds to a build up of fast-moving small-scale structure.

\begin{figure}
\centerline{\psfig{file=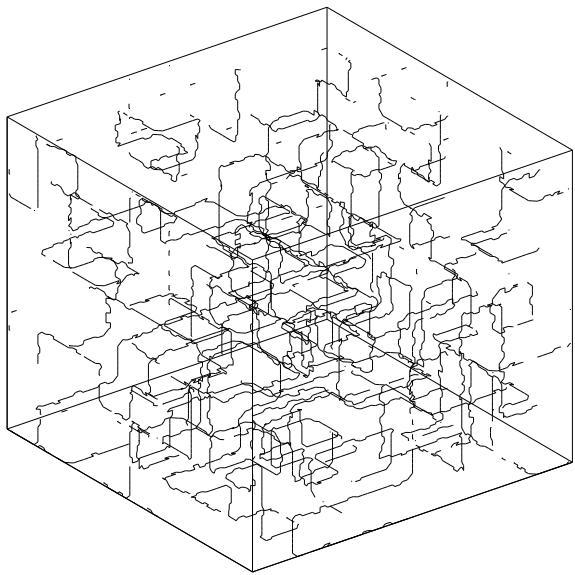,width=3.0in,angle=0}}
\caption{Strings formed with the Vachaspati-Vilenkin algorithm
in a $(128\delta)^3$ box with $\xi_0=8\delta$.
This snapshot is taken shortly after formation.}
\label{fig:VVbeg}
\end{figure}

\begin{figure}
\centerline{\psfig{file=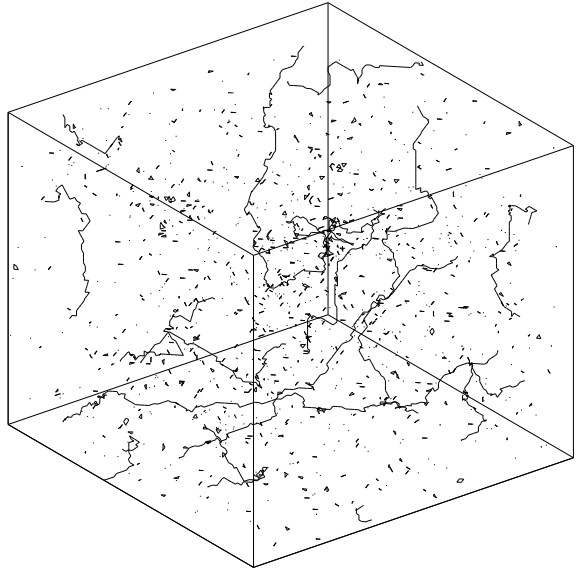,width=3.0in,angle=0}}
\caption{The strings pictured 
in the previous figure after the evolution period
with $E_c=4\delta$.}
\label{fig:VVend}
\end{figure}

Initial string configurations are generated using
the Vachaspati-Vilenkin algorithm \cite{Lattice}, which mimics the
breaking of a $U(1)$ symmetry during a cosmological
phase transition. In this approach, phases from the minimally
discretised $U(1)$ group are placed randomly on the sites of a cubic
lattice. Strings (or anti-strings) are identified as passing through
a face with a net winding about the manifold. The strings are then 
joined up
within the lattice cells to from a network (any ambiguity
if two strings leave and two strings enter a lattice cell is settled
randomly). The domains, and consequently
the initial network step size
$\xi_0$, are of constant size. This initial configuration
is defined on a cubic lattice with fundamental lattice spacing 
$\delta$.
We can alter the initial correlation length $\xi_0$ in terms of 
$\delta$.
If each initial segment length is made up of a large number of 
individual
stationary string links, peculiarities arise in the scaling 
solution as entire segments can be annihilated into 
loops of energy $E_c$ moving at the speed of light.
Consequently
we add structure in the form of cusps to break up long straight 
segments. Cusps
are string links confined to one lattice site which move at the 
speed of light.
Figures \ref{fig:VVbeg} and \ref{fig:VVend} show a 
Vachaspati-Vilenkin string
network shortly after formation and after a period of evolution.
\begin{figure}
\centerline{\psfig{file=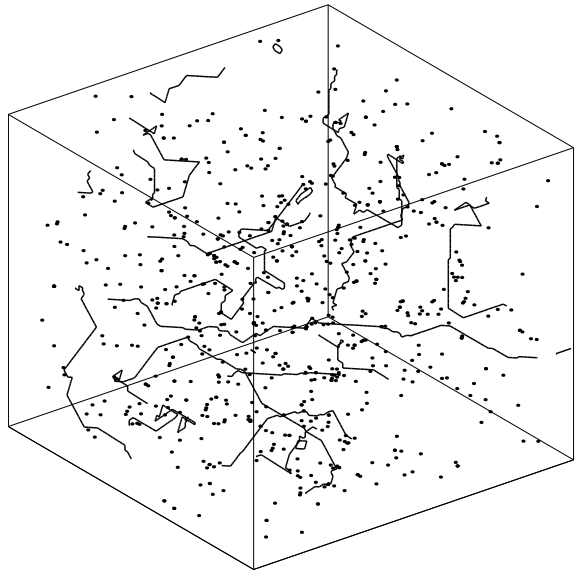,width=3.0in,angle=0}}
\caption{The strings pictured 
in Figure \ref{fig:VVbeg} after the evolution period
with $E_c=2\delta$. The number of small loops is suppressed.}
\label{fig:VVend2}
\end{figure}
We have studied simulations on lattices ranging from $(64\delta)^3$
to $(1024\delta)^3$ . We varied $\xi_0$ from $8\delta$ to $64\delta$ 
and $E_c$ from
$2\delta$ to $8\delta$. Periodic boundary conditions are imposed 
throughout.
We restrict the evolution time to half the box size, as after this 
time causal
influences have propagated around the box.

An ensemble typically consists of 30-50 runs. Unless otherwise
stated, measurements were taken from networks created 
in a $(128\delta)^3$ box with $\xi_0=8\delta$.

\section {Left and right movers}

The most general solution to Eq. (\ref {Waveequation}) for cosmic 
strings
in a Minkowski spacetime is
\begin{equation}
{\bf X}({s},\eta)={1\over 2 }[{\bf a}({s}-\eta)+{\bf b}({s}+\eta)].
\label{GenSol}
\end{equation}
This solution may be considered as made up of ``left-moving''
(${\bf a}$) and ``right-moving'' (${\bf b}$) pieces.
We will consider the dynamical quatities 
${\bf p}$ and ${\bf q}$ defined 
\begin{equation}
{\bf p} = {\bf b'} = \dot {\bf X} + {\bf X}',    
\label{defp}
\end{equation}
\begin{equation}   
{\bf q} = - {\bf a'} = \dot {\bf X} - {\bf X}',  
\label{defq}
\end{equation}
where ${\bf p}^2={\bf q}^2=1$. The solution (\ref{GenSol})
may then be represented by a pair of curves for ${\bf p}$ and 
${\bf q}$ on a unit (``Kibble-Turok'') sphere.
String intercommutation can excise a loop
between ${s}_1$ and ${s}_2$ if 
\begin{equation}  
\int_{{s}_1}^{{s}_2}  d{s} ({\bf p}-{\bf q}) =0.
\label{looprodcond}
\end{equation}
At such an intercommutation event, kinks are created as abrupt 
changes
in the ${\bf p}$ and ${\bf q}$ vectors. Left and right moving kinks 
travel away from
the intercommutation site in opposite directions. Correlations 
between
${\bf p}$s and ${\bf q}$s along the string are extremely 
important \cite{ACK,Hind93}
as they relate to loop production, but cannot easily be calculated 
{\em a priori}.
If ${s}={s}_1-{s}_2$ 
\begin{equation}  
C_{pp}(s,\eta') = \langle {{\bf p}(s_1)\cdot {\bf p}(s_2)\rangle}
\vert_{\eta=\eta'},
\label{cpp}
\end{equation}
\begin{equation}  
C_{pq}(s,\eta') = \langle {{\bf p}(s_1)\cdot {\bf q}(s_2)\rangle}
\vert_{\eta=\eta'},
\label{cpq}
\end{equation}
These two correlation functions contain all the information needed 
as
$C_{qq}(s)=C_{pp}(s)$ and $C_{qp}(s)=C_{pq}(-s)$.
We find that these functions approximately
scale with $\xi$ i.e. they are functions of 
$s/\xi$ only. We calculated these correlation functions
over an ensemble of 50 realisations for various $E_c$.
The results for the scaling
functions for $E_c=2\delta$ are plotted in Figure (\ref{fig:ppVV}). 
The ensemble averages are fitted well by the simple functions:
\begin{figure}
\centerline{\psfig{file=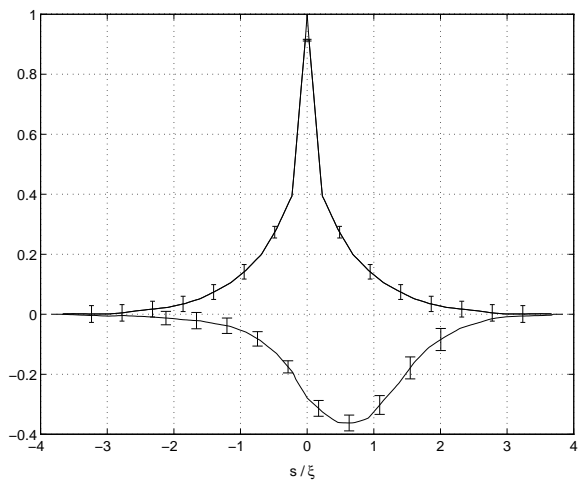,width=3.0in,angle=0}}
\caption{Scaling $C_{pp}$ (upper curve) and $C_{pq}$ correlation 
functions
for $\xi_0=8\delta$ and $E_c=2\delta$.}
\label{fig:ppVV}
\end{figure} 
\begin{equation}
C_{pp}(|s|/\xi) = (1-\omega)e^{-a|s|/\xi} + \omega e^{-b|s|/\xi},
\label{toy}
\end{equation}
\begin{equation}
C_{pq}(s/\xi) = de^{-f (s/\xi-e)^2 }.
\label{gauss}
\end{equation}
The first of these was used in \cite{ACK} for demonstration
purposes, but is in fact a good approximation to our measurements.
It is particularly good for the $E_c=2\delta$ case. The gaussian
form for $C_{pq}$ becomes less good away from the peak
as the tail may become exponential. The gaussian fit is however
within the ensemble errors.
\begin{table}[ht]
\begin{center}
\begin{tabular} {|c|c|c|c|}
\hline
parameter & $E_c=2\delta$ &  $E_c=4\delta$ & $E_c=8\delta$\\
\hline
$\omega$ & 0.57 $\pm$ 0.05 &  0.45 $\pm$ 0.03 & 0.25 $\pm$ 0.03\\
$a$ & 5.4 $\pm$ 1.2  & 20 $\pm$ 3 & - \\
$b$ & 0.54 $\pm$ 0.1 & 0.35 $\pm$ 0.04 &  0.17 $\pm$ 0.03 \\
$d$ & -0.34 $\pm$ 0.02 & -0.20 $\pm$ 0.02& -0.08$\pm$ 0.02\\
$f$ & 0.55 $\pm$ 0.1 & 0.30 $\pm$ 0.05 & 0.03 $\pm$ 0.01\\
$e$ & 0.70 $\pm$ 0.1 &1.1 $\pm$ 0.1 & 3.8 $\pm$ 0.2 \\
\hline
\end{tabular}
\caption{Parameters for model $C_{pp}$ and $C_{pq}$ functions.}
\label{tab:modelparams}
\end{center}
\end{table}
The parameters for the model functions are given in Table 
\ref{tab:modelparams}.
The parameters $a$ and $b$ pick out two length scales along the 
string.
The effect of increasing $E_c$, and therefore small scale structure, 
is to decrease
the small scale $1/a$ and increase the large scale $1/b$, relative 
to $\xi$.
However, for $E_c>4\delta$ the small scale set by $1/a$ is no longer 
constant
and $C_{pp}$ and $C_{pq}$ do not truly scale.

\begin{figure}
\centerline{\psfig{file=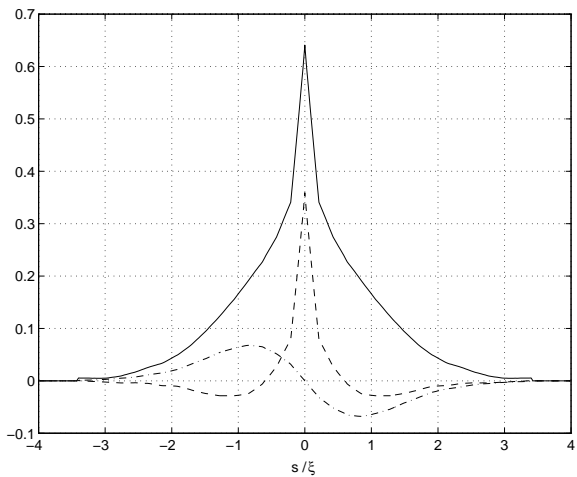,width=3.0in,angle=0}}
\caption{The scaling functions for correlations between the 
following
vectors: tangent-tangent ($-$), velocity-velocity ($-\,-\,$) and
tangent-velocity ($-\,\cdot$) for $\xi_0=8\delta$ and 
$E_c=2\delta$.}
\label{fig:deriv_correl}
\end{figure}

The related correlations between the tangent and velocity vectors
along the string are shown
in Figure \ref{fig:deriv_correl}.
These functions can be calculated as simple sums of (\ref{toy})
and (\ref{gauss}) using equations (\ref{defp}) and (\ref{defq}).

\section {Length scales and rate equations}
Analytic work as so far led to three dynamical length scales 
characterising the network.
Austin, Copeland and Kibble \cite {ACK} have considered network 
evolution
in terms of the dynamics of left- and right-moving kinks and have 
defined three
length scales used in their analysis which we can calculate using 
our simulations:

(i) $\xi$, the familiar energy density length scale

(ii) $\bar\xi$, a persistence length along the string

(iii)  $\zeta$, a measure of the small-scale kinky structure.
Their definitions are 
\begin{eqnarray}
\xi(\eta)& =& \sqrt{\mu/\rho_l(\eta)},\\
\bar\xi(\eta)& = &\int\limits_0^{\infty} ds \langle{ {\bf 
p}(s_1).{\bf p}(s_2)\rangle},\\
& = & ((1-\omega)/a + \omega/b)\xi,\\
\zeta^{-1}(\eta)& = & -1/{\partial \over \partial s} \langle {{\bf 
p}(0).{\bf p}(s)\rangle}  \vert_{s=0},\\
& = & ((1-\omega)a + \omega b)\xi^{-1},
\label{lengths}
\end{eqnarray}
where $\xi$ and $\bar\xi$ are also expressed in terms of the
specific correlation function given in (\ref{toy}).

The angular brackets $\langle{ \rangle}$ indicate averaging over the 
long string, which is defined
as all string having energy greater than $\xi$, {\em and} over an 
ensemble of 
realisations.

ACK developed rate equations for the three length scales. From their 
(expanding universe)
analysis it seemed most likely that the network would enter a 
transient scaling regime, in
which $\xi$ and $\bar\xi$ scaled but $\zeta$ did not. ($\zeta$ would 
scale 
only after gravitational back reaction was taken into account, when 
${\zeta / \bar\xi} \approx 10^{-4}$) .
In order to study this solution, we considered
these rate equations and made an approximation
suitable to our simulations: we ignore terms involving Hubble 
expansion
or gravitational radiation. Then the rate equations become:
\begin{equation}
{\dot\xi \over \xi} = {c \over 2\bar\xi},
\label{rate1}
\end{equation}
\begin{equation}
{\dot{\bar\xi} \over \bar\xi} = { -\chi \bar\xi \over \omega \xi^2} 
+ { I \over 2 \bar\xi },
\label{rate2}
\end{equation}
\begin{equation}
{\dot\zeta \over \zeta} = { -\chi \zeta \over \xi^2 } + { kc \over 
\bar\xi }.
\label{rate3}
\end{equation}
In Eq. (\ref{rate1}), $c$ characterises loop production.
ACK consider it as function of ratios of length scales, although
it should tend to a constant in the scaling regime of our 
simulations.
In Eq. (\ref{rate2}), $\chi$ is a geometrical factor 
affecting the frequency of intercommutation between uncorrelated
string segments; $\omega$ is a parameter in the correlation function 
$C_{pp}$ as described by Eq. (\ref{toy}). The function $I$
also relates to loop production. In Eq. (\ref{rate3}), 
$k$ relates to the efficiency of small-scale structure removing 
itself
from the long string network.
We change variables to $x = \xi /\eta$, $\bar x = \bar\xi/\eta$ and 
$z = \zeta/\eta$ to express
the above rate equations:
\begin{equation}
{\eta \dot x} = {c x \over 2 \bar x} - x,
\label{ratex}
\end{equation}
\begin{equation}
{\eta \dot{\bar x} } = { -\chi \bar x^2 \over \omega x^2} + { I 
\over 2} - \bar x,
\label{ratexb}
\end{equation}
\begin{equation}
{\eta \dot z} = { -\chi z^2 \over x^2 } + { k c z \over \bar x}-z.
\label{ratez}
\end{equation}

\section{The Approach to Scaling}

As in ACK we express each length scale 
as the product of time and a scaling variable so that $\xi=x\eta$, 
$\bar\xi=\bar x\eta$ and
$\zeta=z\eta$. It has been widely reported \cite 
{AlbTur8589,FRWCodes,SakVil90} that $\xi$ scales.
\begin{figure}
\centerline{\psfig{file=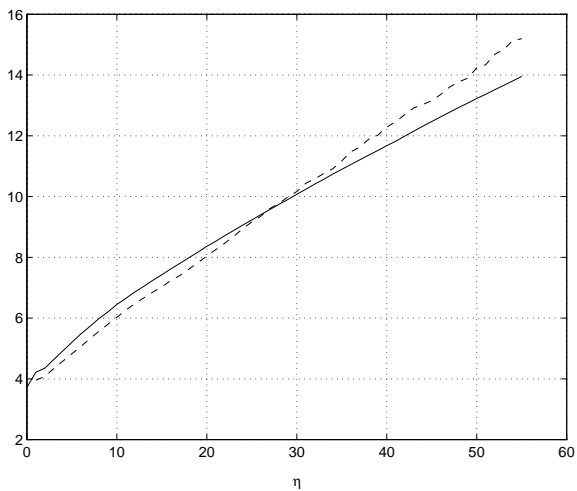,width=3.0in,angle=0}}
\caption{Plot of $\xi$ and $\bar\xi$ for a $(128\delta)^3$ box with
$\xi_0=4\delta$ and $E_c=2\delta$.}
\label{fig:xibarxi}
\end{figure} 
Inspection of Figure \ref{fig:xibarxi} indicates the possibility 
that 
$\xi$ and  $\bar\xi$ are not in fact scaling, but growing with a 
power of $\eta$ less than 1. A correction to the standard
scaling scenario is also suggested by a recent analytical 
calculation
\cite{Hind96}.  A study of the exponent $n$ in $\eta^n$ as the 
network
evolves shows that it is indeed tending towards $1$ throughout the 
simulations.
At late times we find no significant departure from 
the scaling value $n=1$. For a $(128\delta)^3$ box, 
$n=0.96\pm 0.05$ which may indicate than $\xi$ is still on an 
approach to scaling.
However, as the box size is increased, $n$ becomes ever closer to 1
with $n=1.000\pm 0.003$ for a $(1024\delta)^3$ box.
The errors on $\bar\xi$ are larger, but similar conclusions 
apply.
To extract the scaling values for the two length scales,
we fit the data $\xi/\eta$ towards the end of the simulation to 
$x_* + \xi_*/\eta$, where $\xi_*$ and $x_*$ are free. This is
motivated in part by the approach to scaling predicted from the ACK 
model
at the end of this section.

The resulting scaling values $x_*$ and $\bar x_*$
are given in Tables \ref{tab:scalx} and \ref{tab:scalxb}.
The figures are given with 1-sigma error bars from averaging over an 
ensemble
of realisations. 
\begin{table}[ht]
\begin{center}
\begin{tabular} {|c|c|c|c|}
\hline
$\xi_0$ & $E_c=2\delta$ &  $E_c=4\delta$ & $E_c=8\delta$\\
\hline
$8\delta$ & 0.174 $\pm$ 0.018 & 0.130 $\pm$ 0.012 & 0.089 $\pm$ 
0.006\\
$16\delta$ & 0.170 $\pm$ 0.010 & 0.141 $\pm$ 0.010 & 0.109 $\pm$ 
0.004 \\
$32\delta$ & 0.171 $\pm$ 0.011 & 0.147 $\pm$ 0.006 & 0.117 $\pm$ 
0.006 \\
$64\delta$ & 0.173 $\pm$ 0.011 & 0.144 $\pm$ 0.003 &  $0.122\pm 
0.014$\\
\hline
\end{tabular}
\caption{Scaling values for $x=\partial\xi/\partial\eta$.}
\label{tab:scalx}
\end{center}
\end{table}

\begin{table}[ht]
\begin{center}
\begin{tabular} {|c|c|c|c|}
\hline
$\xi_0$ & $E_c=2\delta$ & $E_c=4\delta$  & $E_c=8\delta$ \\
\hline
$8\delta$ & 0.19 $\pm$ 0.04 & 0.13 $\pm$ 0.03 & 0.070 $\pm$ 0.03\\
$16\delta$ & 0.20  $\pm$ 0.04 & 0.17 $\pm$ 0.03 & 0.10 $\pm$ 0.03\\
$32\delta$ & 0.19 $\pm$ 0.03 & 0.17 $\pm$ 0.03 & 0.12 $\pm$ 0.04\\
$64\delta$ & 0.25 $\pm$ 0.04 & 0.21 $\pm$ 0.04 & $0.14\pm 0.03$\\
\hline
\end{tabular}
\caption{Scaling values for $\bar x=\partial\bar\xi/\partial\eta$.}
\label{tab:scalxb}
\end{center}
\end{table}
The scaling parameter $x_*$ is independent of $\xi_0$,
within ensemble errors. It is however strongly dependent 
on $E_c$ to the extent that, with an increase of $E_c$ from
$2\delta$ to $8\delta$, the scaling density increases by a factor of 
2.
The scaling values of $\bar x_*$ have bigger ensemble errors,
and this may hide some dependence on $\xi_0$. However, the results
so far show that, within the ensemble errors the only dependence
is on $E_c$.

\begin{figure}
\centerline{\psfig{file=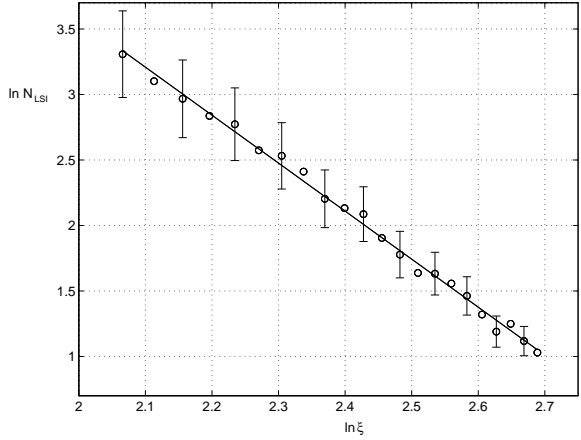,width=3.0in,angle=0}}
\caption{The number of long string intercommutation events 
($N_{LSI}$)
in the simulation box at each timestep plotted against $\xi$.}
\label{fig:Ntime}
\end{figure}

One of the problems testing the model is the difficulty
in calculating the parameters directly. We must use the model to
calculate some of the parameters and see whether this gives
consistent results.
Assuming that $c$, $I$, $\chi$ and $k$
achieve scaling values then the following expressions are
stable fixed points.
\begin{equation}
\bar x_* = { c \over 2 },
\label{xbscales}
\end{equation}
\begin{equation}
x_* = \sqrt{ {\chi c^2 \over 2 \omega (I - c)}},
\label{xscales}
\end{equation}
\begin{equation}
z_* = \left\{ \begin{array}{ll} 
              (2k-1) x_*^2/ \chi & {\rm if} \: 2k-1 > 0\\
              0  & {\rm if} \: 2k-1 \leq 0.\\
             \end{array}
           \right.
\label{zscales}
\end{equation}
We are able to calculate $c$ directly from the scaling value of
$\bar\xi_*$. The value of $\chi$ can be measured by 
counting the number of long string intercommutation events over
each time step. Following ACK, the probability of a segment of 
length 
$\bar\xi$ intercommuting in a time step $\delta$ is given by:
\begin{equation}
{ \chi\bar\xi\delta \over \xi^2 }.
\label{chieq}
\end{equation}
We assume that the number of segments in the volume V
is given by $V/(\xi^2\bar\xi)$, thus the number of long
string intercommutating events in a volume V over $\delta$ is given 
by:
\begin{equation}
{ \chi V \delta \over \xi^4 }.
\label{chieq2}
\end{equation}
Figure \ref{fig:Ntime} shows the number of long string 
intercommutations
as a function of $\xi$ for $E_c=2\delta$, together with a slope 
giving an
$\xi^{-4}$ dependence. From the $y$-intercept we can estimate $\chi$ 
at
$0.05\pm 0.015$ for both $E_c=2\delta$ and $E_c=4\delta$.
We are now able to make a direct test of the rate equations by
considering the approach to scaling of $\xi$.

\begin{figure}
\centerline{\psfig{file=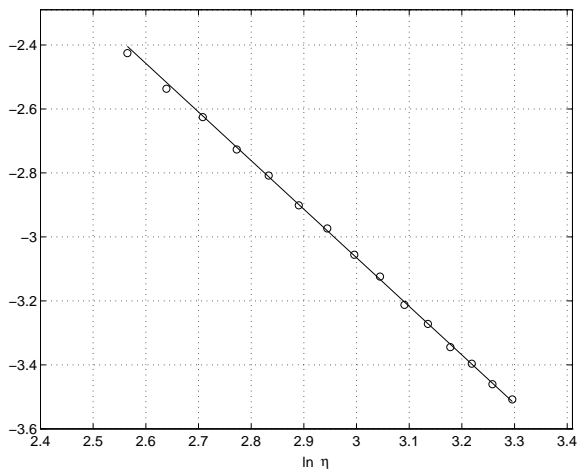,width=3.0in,angle=0}}
\caption{Log plot of the $\eta^{\epsilon}$ term in Eq. 
(\ref{x:app}).} 
\label{fig:second_xi}
\end{figure}

Expanding about the fixed points gives (to first order) a matrix 
equation equivalent to Eqs. (\ref{ratex})-(\ref{ratez}):
\begin{equation}
\eta\dot y = M y,
\label{Meq}
\end{equation}
where
\begin{equation}
 y = \left(\matrix{x-x_*\cr
                         \bar x-\bar x_*\cr
                         z-z_*\cr}\right),
\label{y}
\end{equation}
and 
\begin{equation}
M = \left(\matrix{ {c \over 2 \bar x_*}-1& {- c x_* \over 2 \bar 
x_*^2} & 0\cr
                         {2 \chi \bar x_*^2 \over \omega x_*^3}&
                                        {-2 \chi \bar x_* \over 
\omega x_*^2}-1&0\cr
                         {2 \chi z_*^2 \over x_*^3}&{-k c z_* \over 
\bar x_*^2}&
                           {-2 \chi z_* \over x_*^2}+{k c \over \bar 
x_*}-1\cr}\right).
\label{M1}
\end{equation}
Using the scaling value for $\bar x_*$ and $\bar z_*$ given above M 
reduces to  
\begin{equation}
\left(\matrix{ 0 & {- x_* \over \bar x_*} & 0 \cr
                 {2 \chi \bar x_*^2 \over \omega x_*^3}&{-1-{2 \chi 
\bar x_* \over \omega x_*^2}}&0\cr
                         {2 \chi z_*^2 \over x_*^3}&{-k c z_* \over 
\bar x_*^2}&
                          -|2k-1|\cr}\right). 
\label{M2}
\end{equation}
The eigenvalues are then $-1$, ${-2 \chi \bar x_* \over \omega 
x_*^2}$
and $-|2k-1|$.
Thus we expect in the approach to scaling (for large $\eta$)  the 
time dependences:
\begin{equation}
x(\eta) = C_1 {x_* \over\bar x_* }\eta^{-1} + 
C_2 {\omega x_*^3 \over 2 \chi \bar x_*^2 }\eta^{\epsilon} + x_*,
\label{x:app}
\end{equation}
\begin{equation}  
\bar x(\eta) = C_1 \eta^{-1} + C_2 \eta^{\epsilon} + \bar x_*,
\label{xb:app}
\end{equation}
\begin{equation}  
z(\eta) = C_3 \eta^{\lambda} + z_*,
\label{z:app}
\end{equation}
where
\begin{equation}
\epsilon=-2 \chi \bar x_* / \omega x_*^2 ,
\label{epsdef}
\end{equation}
\begin{equation}
\lambda=-|{2k-1}|.
\label{lamdef}
\end{equation}
To compare with the simulations, we measured $x=\xi/\eta$. 
As predicted from (\ref{x:app}), we measured
a $\eta^{-1}$ dependence for large $\eta$, although this
arises partly from $\xi_0/\eta$.
We subtract the $x_*+a\eta^{-1}$ dependence from a plot of $x$, 
leaving
a term proportional to $\eta^{\epsilon}$. The resulting log-log plot 
for a simulation with 
$\xi_0=4\delta$ and $E_c=2\delta$ is shown in Figure 
(\ref{fig:second_xi}),
and it gives a good single power law fit to $\epsilon={-1.4}$.
The ACK model predicts $\epsilon=-1.5\pm 0.2$,  from Eq. 
(\ref{epsdef}) and our measured values for
$\bar x_*$, $x_*$ and $\omega$. Similarly for $E_c=4\delta$
we measure $\epsilon=-1.8\pm 0.2$ and the model predicts 
$\epsilon=-1.7\pm 0.2$.
We see no evidence that another scale enters into the approach to 
scaling.

\section{Small Scale Structure}

\begin{figure}
\centerline{\psfig{file=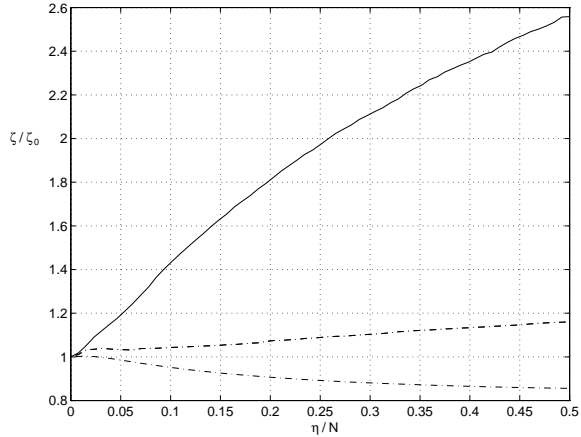,width=3.0in,angle=0}}
\caption{The behaviour of $\zeta$ for $E_c=2\delta$ ($-$), 
$E_c=4\delta$ ($--$)
and $E_c=8\delta$ ($-\cdot$). The plots correspond to different box 
sizes to
ensure that $\xi_0/E_c \geq 4$.}
\label{fig:zetasp}
\end{figure}

Austin {\em et\, al} characterise small-scale structure using 
${\bf p}$ and ${\bf q}$ vectors to define a length scale,
as in Eq. (\ref{lengths}), from which we get:
\begin{equation}
\zeta^{-1}(\eta) = \langle{ {\bf p}(0) \cdot {\bf p'}(0)\rangle}.
\label{zeta2}
\end{equation}
Figure \ref{fig:zetasp} shows the dependence of $\zeta$ on $E_c$.
An $E_c$ of $2\delta$ allows structure to rapidly leave the network
and $\zeta$ to grow. For $E_c=8\delta$, structure builds
up on the string faster then loop production can remove it and 
$\zeta$
decays. 
For $E_c=2\delta$ and $E_c=4\delta$, $\zeta$ appears to
approach a scaling value. For a box of size $(128\delta)^3$,
the exponent $n$ in $\zeta=\zeta_* + z\eta^n$ achieves $n=0.98 \pm 
0.03$ 
towards the end of the simulation and $z_*$ tends towards the values 
given 
in Table \ref{tab:scalez}, given for a range of string densities.
We note that for both these values of $E_c$,
$\zeta/\xi\gg 10^{-4}$: much bigger than the ratio
required for gravitational radiation to become significant.
Indeed,
for the smallest value, $\zeta$ scales at the same magnitude as 
$\xi$.
As we increase $E_c$ however, we find that $\zeta$ tends towards 
zero growth. 
This signifies the build up of small-scale structure which would 
eventually 
trigger effective gravitational back-reaction. We would like to 
encompass the
dependence of $\zeta$ on $E_c$ in terms of the three-scale model.
\begin{table}[ht]
\begin{center}
\begin{tabular} {|c|c|c|}
\hline
$\xi_0$ & $E_c=2\delta$ & $E_c=4\delta$\\
\hline
$8\delta$   & $0.11\pm 0.01$   &  $0.007\pm 0.002$\\
$16\delta$   & $0.12\pm 0.01$  & $0.008\pm 0.002$ \\
$32\delta$    & $0.12\pm 0.01$  & $0.005\pm 0.001$ \\
$64\delta$    & $0.12\pm 0.01$  & $0.005\pm 0.002$ \\
\hline
\end{tabular}
\caption{Scaling values for $z_*=\partial\zeta/\partial\eta$.}
\label{tab:scalez}
\end{center}
\end{table}
The ACK model for $\zeta$, Eq. (\ref{rate3}), consists of two 
processes.
The first is the effect of long string intercommutation. This will
serve to create kinky structure
and tends to decrease $\zeta$.  The second term involves the effect 
of small
loop production.  It is controlled by a parameter $k$ which reflects
the extent to which loop production can remove small scale 
structure.
If we assume a scaling regime for $\xi$ and $\bar\xi$
we can integrate Eq. (\ref{ratez}) to get
\begin{equation}  
z(\eta) = { (2k-1)\eta^{2k-1} \over {C+ \chi \eta^{2k-1} x_*^{-2} }},
\label{z:exac}
\end{equation}
where $C$ is a constant of integration.
The evolution of $\zeta$ is then crucially dependent on
$k$. For $k>\frac{1}{2}$, $\zeta$ has a positive gradient and $z$ 
tends towards the scaling value
$z_*=(2k-1)x_*^2/\chi$. For $k<\frac{1}{2}$, $\zeta$ has a negative 
gradient and tends towards 0.
From Figure \ref{fig:zetasp} we see that both cases occur in our 
simulations,
depending on the value of $E_c$. The three-scale model is certainly 
capable of explaining
our simulations, if we consider $E_c$ as being related to $k$. This
should not surprise us, as increasing $E_c$ artificially
keeps structure on the string that would otherwise leave the 
network.
Unfortunately, we cannot calculate $k$  {\em a priori}. We can only
assume that the rate equations
from the three-scale model are broadly correct  and
extract a 
value for 
$k$ numerically, by using the scaling expression for $z_*$. 
This gives $k=0.59$ ($E_c=2\delta$), $k=0.51$ ($E_c=4\delta$)
and  $k<\frac{1}{2}$ ($E_c>6\delta$). 

\begin{figure}
\centerline{\psfig{file=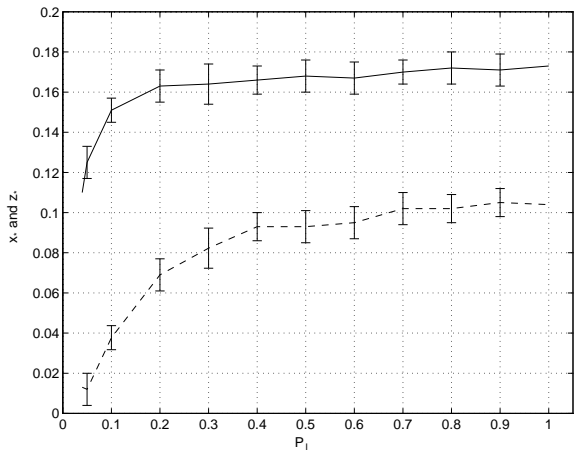,width=3.0in,angle=0}}
\caption{How the scaling values $x_*$ and $z_*$ vary with
the intercommutating probability.}
\label{fig:xz_with_prob}
\end{figure}
The effect of changing the intercommutating probability on $z_*$
is broadly similar to the effect of $x_*$ \cite{SakVil90}. We 
compare 
the two in Figure \ref{fig:xz_with_prob}. The value for $z_*$ 
changes
slowly until $P_I\approx 0.35$ when it falls rapidly. After
this point $\zeta$ is probably not scaling.

Another way to characterise small-scale structure along
the string is through the relationship
\begin{equation}
E=R^d,
\label{FractRel}
\end{equation}
where $E$ is the average energy between two points separated
by a physical distance $R$. The exponent
$d$ is scale dependent.  On large scales it is shown in 
\cite{FRWCodes,SakVil90} that $d$ tends to $2$, as the strings become
random walks. On small-scales, the string becomes straight on 
average,
and $d$ tends to 1.  One might expect a smooth interpolation between
these two regimes, but Bennet and Bouchet \cite{FRWCodes} note that 
an
``intermediate'' fractal $d_i$ (i.e. a constant or very slowly 
varying $d$)
exists on scales between the numerical cutoff and below $\xi$.
They explain this as the  build up of small scale structure on
over the scales of constant fractal.

\begin{figure}
\centerline{\psfig{file=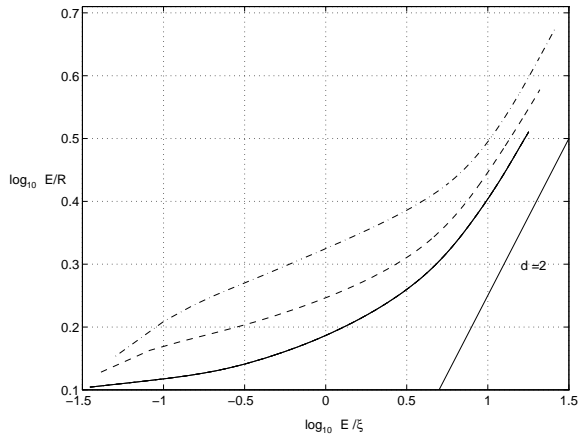,width=3.0in,angle=0}}
\caption{Log plot showing the fractal structure of the long string
for energy cutoffs of $E_c=2\delta$ (---), $E_c=4\delta$ (-\,-)
and $E_c=8\delta$ (-$\cdot$).}
\label{fig:ln_fract}    
\end{figure}

\begin{figure}
\centerline{\psfig{file=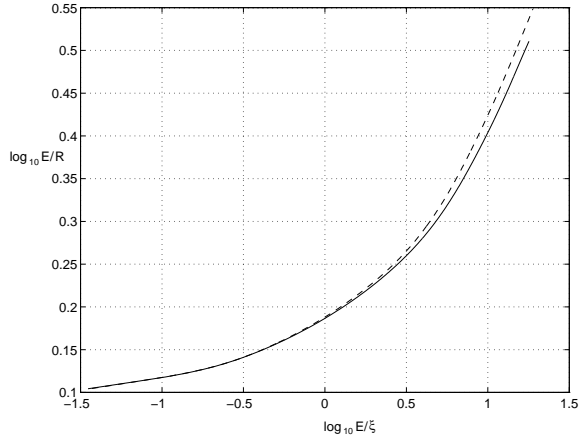,width=3.0in,angle=0}}
\caption{Log plot showing the fractal structure with a cutoff of 
$E_c=2\delta$ (---) together with the same quantity calculated
from the model tangent-tangent correlation function.}
\label{fig:ln_fract_tantan}    
\end{figure}

Allen and Shellard \cite{FRWCodes} observe a similar structure, 
although they
attribute it to artificial initial conditions and lack of dynamical 
range.
We observe that this intermediate fractal exists for energy cut-offs
of $8\delta$ and above whereas for $E_c=2\delta$ it has
disappeared completely leaving a smooth transition from $d=1$ to 
$d=2$.
This is shown in Figure \ref{fig:ln_fract}, which is plotted for
comparison with similar figures in \cite{FRWCodes}. 
However, it is not clear that small scale structure always
reveals itself through a constant $d_s$. 
As a comparison, we calculated $E/R$ from the tangent-tangent
correlation function $\langle{{\bf X'}(z)\cdot{\bf X'}(0)\rangle}$
with the parameters in Table \ref{tab:modelparams} for 
$E_c=2\delta$.
For these model parameters, we find that $\zeta\approx 0.4\xi$:
certainly not a small scale.
We plot the result in Figure \ref{fig:ln_fract_tantan} along with 
the measured quantity.
The agreement is good up to scales well in excess of $\xi$. (The 
discrepancy 
at larger scales may be due in part to model function
underestimating the ${\bf pq}$ correlations at large
$\xi$, as the gaussian form of $C_{pq}$ breaks down.)
However, for small scales the agreement is good and this indicates
a genuine lack of small-scale structure.

\section{Conclusions}

We have demonstrated the scaling behaviour of correlation functions
of ${\bf p}$ and ${\bf q}$ vectors, which are essential in analytic 
studies of string
networks. 

The parameters of model functions for the ${\bf pp}$ and ${\bf qq}$
correlation functions $C_{pp}$ and $C_{pq}$, depend on how much
small-scale structure in present on the string, which is related
to the value of the minimum allowed
loop energy $E_c$. As $E_c$ is increased, small-scale structure
is allowed to build up and the scaling density can increase by a
factor of two more.

We analysed our simulations in terms of the three-scale model
proposed by Austin {\it et al}, suitably simplified for Minkowski
space with no gravitational radiation. 

The model for $\xi$ describes well the approach to scaling and 
predicts
the two exponents for this approach that we measure in our 
simulations.
The model for the ``small'' scale $\zeta$ is qualitatively 
consistent
with our simulations with critical value $k$ determining whether or
not $\zeta$ will scale. 

For the smallest possible value of $E_c$, small-scale structure
disappears, as all the scales $\xi$, $\bar\xi$ and $\zeta$ 
of the 3-scale model of ACK have the same order of magnitude.
Furthermore, from the fractal analysis along the string for 
$E_c=2\delta$, the transient fractal
region disappears, leaving a smooth transition from straight
strings on small-scales to the random walk on large scales.
Thus the intermediate fractal region seen in the expanding Universe 
codes
seems to be an effect of the limit on the size of loops.

In all cases, loops are predominantly produced at the smallest
possible scale $E_c$, even when it is set, by the underlying 
lattice,
to $2\delta$. Thus, in common with all other simulations (apart from 
those of Albrecht and Turok \cite{AlbTur8589}),
there is no evidence for scaling
in loop production. It is therefore time to ask the question: is
this the true physical situation? It is certainly possible that 
loops are produced with the smallest possible physical scale, which 
is the 
string width, and we see from our simulations that every other 
measure of 
the network has a single scale of order $\xi$. Thus there is no real
conflict with the string scaling hypothesis.

One objection to the use of the Smith-Vilenkin
algorithm is its restrictive set of possible values
of ${\bf p}$ and ${\bf q}$ for the initial conditions
(although the subsequent evolution is exact). As discussed
by Albrecht \cite{Alb90}, this may exaggerate small loop production.
However, expanding Universe codes have no such restrictions and 
there is no published
evidence that loop production happens at any scale other than the 
loop size cut-off.
Furthermore, we have found no evidence that the lattice scale enters 
the 
scaling solution. If there is spurious cusp
production (for $E_c=2\delta$) through a chance configuration 
of ${\bf p}$ and ${\bf q}$ vectors such that Eq. (\ref{looprodcond}) 
is true,
then one would not expect a stable scaling solution. 
Albrecht suggests that there may also be spurious
loop production through ``back-tracking'' \cite{Alb90} which occurs 
when individual discretised links or groups of links point in a 
direction opposite  
to that of the whole segment of length $\xi$. In this case, one 
would expect
the lattice scale $\delta$ to affect the evolution: for example, the
rate of energy loss from back-tracking would most likely increase
with $\xi/\delta$ as the chances for a given link
or group of links to back-track in a time $\xi$ increases.

We are confident that our simulations {\it do} approach (and for 
large
simulations reach) a stable scaling solution, 
and from Table \ref{tab:scalx} we see that the value of the scaling
parameter $x_*$ is independent of the range of $\xi$ over which 
the simulation is run. We infer from this that the spurious 
processes
mentioned above do not critically enter the observed dynamics.

The prospect that the loops are produced with such tiny sizes is a 
radical one.
It means that the dominant mode of energy loss of a cosmic string 
network is
particle production and not gravitational radiation as the loops 
collapse almost
immediately. Indeed, it may not be possible to talk about loop 
production at all:
the loop production we observe favours loops formed at cusps,
which could annihilate into particles before the loop is formed 
\cite{Srednicki87}.
Furthermore, gravitational radiation directly from a network without
a small scale $\zeta\ll\xi$ is  negligible.

Recent calculations \cite{Gill94} on the possible role of
GUT quanta decaying into high energy cosmic rays assumes that 
particle
production is {\em sub}dominant to gravitational radiation
as a means of energy loss, and conclude that GUT strings could not 
be an
appreciable source of cosmic rays.
In our suggested energy loss scenario, we may come up against 
an observational bound of high energy cosmic rays. Simple models
of GUT quanta decay processes give the upper bound for the 
proportion
of the string energy going into high energy cosmic rays as 
$O(10^{-3})$
or less \cite{PartBound}. Such calculations may put a significant
constraint on GUT models if the GUT scale string scenario is to be 
viable.

\section*{Acknowledgements}

We wish to thank Ed Copeland for useful discussions.

GRV and MBH are supported by PPARC, by studentship number 
94313367,  Advanced Fellowship number B/93/AF/1642 
and grant number GR/K55967. MS is 
supported by the Tomalla Foundation. Partial support is
also obtained from the European Commission under the Human Capital 
and Mobility programme, contract no. CHRX-CT94-0423.


\begin{thebibliography} {99}

\bibitem{HindKib94} M. Hindmarsh and T.W.B. Kibble 
{\em Rep. Prog. Phys.} {\bf 58} 477 (1994)
\bibitem{ShelVil} A. Vilenkin and E.P.S. Shellard,
{\em Cosmic Strings and other Topological Defects}
(Cambridge University Press, Cambridge, 1994)
\bibitem{Lattice} T. Vachaspati and A. Vilenkin {\em Phys.  Rev.} 
{\bf D30}, 2036 (1984);
T. W. B. Kibble {\em Phys.  Lett.} {\bf 166B}, 311 (1986);
K. Strobl and M. Hindmarsh to be published in {\em Phys.  Rev.} {\bf 
E55} (1997);
A recent lattice-free calculation has cast some doubt on 
the presence of infinite string:
J. Borrill {\em Phys.  Rev. Lett.} {\bf 76} 3255 (1996) 
\bibitem{Kibble85}  T.W.B. Kibble 
{\em Nucl. Phys.} {\bf B252} 277 (1985)
\bibitem {AlbTur8589} A. Albrecht and N. Turok {\em Phys.  Rev. 
Lett.}{\bf 54}, 1868 (1985);
A. Albrecht and N. Turok {\em Phys.  Rev.} {\bf 
D40}, 973 (1989)
\bibitem{FRWCodes} D. P. Bennett, in ``Formation and Evolution of 
Cosmic Strings'',
eds. G. Gibbons, S. Hawking and T. Vachaspati, (Cambridge 
University Press, Cambridge. 1990); F. R. Bouchet {\it ibid.};
E. P. S. Shellard and B. Allen  {\it ibid.}
\bibitem{SakVil90} M. Sakellariadou and A. Vilenkin {\em Phys. Rev.} 
{\bf D42} 349 (1990)
\bibitem{ACK} D. Austin, E. J. Copeland and T. W. B. Kibble {\em 
Phys. Rev.} {\bf D48} 5594 (1993)
\bibitem{GravInf}
M. Hindmarsh {\em Phys.  Lett.} {\bf 251B} 28 (1990)
M. Sakellariadou  {\em Phys. Rev.} {\bf D42} 354 (1990)
\bibitem{Hind96} M. Hindmarsh {\em Phys. Rev. Lett.} {\bf 77} 4495 
(1996)
\bibitem{Martins96} C. J. A. P. Martins and E. P. S. Shellard {\em 
Phys. Rev.} {\bf D54} 2535 (1996)
\bibitem{VHS} G. R. Vincent, M. Hindmarsh and M. Sakellariadou, to 
be published in
{\em Phys.  Rev.} {\bf D55}, (1997) 
\bibitem {Hind93} M. Hindmarsh  {\em Ap. J.} {\bf 431} 534 (1994) 
\bibitem{Coul94} D. Coulson, P. Ferreira, P. Graham and N. Turok
{\em Nature} {\bf 368} (1994)
\bibitem{Srednicki87} M. Srednicki and S. Theisen {\em Phys. Rev.} 
{\bf B189} 397 (1987)
\bibitem{PartBound} P. Bhattacharjee and N. C. Rana {\em Phys. 
Lett.} {\bf 246B} 365 (1990);
G. Sigl, K. Jedamzik, D. N. Schramm and V. Berezinsky {\em Phys. 
Rev.} {\bf D52} 6682 (1995)
\bibitem{Gill94} A. J. Gill and T. W. B. Kibble {\em Phys. Rev.} 
{\bf D50} 3660 (1994)
\bibitem{SakVil88} M. Sakellariadou and A. Vilenkin {\em Phys. Rev.} 
{\bf D37} 885 (1988)
\bibitem{SmithVil:alg} A. G. Smith and A. Vilenkin {\em Phys. Rev.} 
{\bf D36} 990 (1987)
\bibitem {BB} D. P. Bennett and F. R. Bouchet {\em Phys.  Rev. 
Lett.}{\bf 60}, 257 (1988)
\bibitem{Alb90} A. Albrecht, in ``Formation and Evolution of 
Cosmic Strings'',
eds. G. Gibbons, S. Hawking and T. Vachaspati, (Cambridge 
University Press, Cambridge. 1990).
\end {thebibliography}
\end{document}